%Paper: gr-qc/9511080
%From: Renate Loll <loll@odin.aei-potsdam.mpg.de>
%Date: Thu, 30 Nov 1995 11:08:24 +0100 (MET)

%%%%%%%%%%%%%%%%%%%%%%%%%%%%%%%%%%%%%%%%%%%%%%%%%%%%%%%%%%%%%%%%%%%%%
% paper "Making quantum gravity calculable",       %
% by R. Loll, begins here                                           %
%%%%%%%%%%%%%%%%% fonts, definitions,etc.%%%%%%%%%%%%%%%%%%%%%%%%%%%%

\input epsf.tex

\font\rmu=cmr10 scaled\magstephalf
\font\bfu=cmbx10 scaled\magstephalf

\font\it=cmti10 scaled \magstephalf

\rmu

\font\rmus=cmr8
\font\rmuss=cmr6
\font\mait=cmmi10 scaled\magstephalf
\font\maits=cmmi7 scaled\magstephalf
\font\maitss=cmmi7
\font\msyb=cmsy10 scaled\magstephalf
\font\msybs=cmsy8 scaled\magstephalf
\font\msybss=cmsy7
\font\bfus=cmbx7 scaled\magstephalf
\font\bfuss=cmbx7
\font\cmeq=cmex10 scaled\magstephalf

\textfont0=\rmu
\scriptfont0=\rmus
\scriptscriptfont0=\rmuss

\textfont1=\mait
\scriptfont1=\maits
\scriptscriptfont1=\maitss

\textfont2=\msyb
\scriptfont2=\msybs
\scriptscriptfont2=\msybss

\textfont3=\cmeq
\scriptfont3=\cmeq
\scriptscriptfont3=\cmeq

\newfam\bmufam  \textfont\bmufam=\bfu
      \scriptfont\bmufam=\bfus \scriptscriptfont\bmufam=\bfuss

\hsize=15.5cm
\vsize=21cm
\baselineskip=16pt   % Double spacing
\parskip=12pt plus  2pt minus 2pt

\def\b{\beta}
\def\d{\delta}
\def\e{\epsilon}

\def\g{\gamma}

\def\semi{\bigcirc\kern-1em{s}\;}

\def\del{\partial}
\def\ni{\noindent}
\def\R{{\rm I\!R}}

\def\Q{{\mathchoice
{\setbox0=\hbox{$\displaystyle\rm Q$}\hbox{\raise 0.15\ht0\hbox to0pt
{\kern0.4\wd0\vrule height0.8\ht0\hss}\box0}}
{\setbox0=\hbox{$\textstyle\rm Q$}\hbox{\raise 0.15\ht0\hbox to0pt
{\kern0.4\wd0\vrule height0.8\ht0\hss}\box0}}
{\setbox0=\hbox{$\scriptstyle\rm Q$}\hbox{\raise 0.15\ht0\hbox to0pt
{\kern0.4\wd0\vrule height0.7\ht0\hss}\box0}}
{\setbox0=\hbox{$\scriptscriptstyle\rm Q$}\hbox{\raise 0.15\ht0\hbox to0pt
{\kern0.4\wd0\vrule height0.7\ht0\hss}\box0}}}}
\def\C{{\mathchoice
{\setbox0=\hbox{$\displaystyle\rm C$}\hbox{\hbox to0pt
{\kern0.4\wd0\vrule height0.9\ht0\hss}\box0}}
{\setbox0=\hbox{$\textstyle\rm C$}\hbox{\hbox to0pt
{\kern0.4\wd0\vrule height0.9\ht0\hss}\box0}}
{\setbox0=\hbox{$\scriptstyle\rm C$}\hbox{\hbox to0pt
{\kern0.4\wd0\vrule height0.9\ht0\hss}\box0}}
{\setbox0=\hbox{$\scriptscriptstyle\rm C$}\hbox{\hbox to0pt
{\kern0.4\wd0\vrule height0.9\ht0\hss}\box0}}}}

\font\fivesans=cmss10 at 4.61pt
\font\sevensans=cmss10 at 6.81pt
\font\tensans=cmss10
\newfam\sansfam
\textfont\sansfam=\tensans\scriptfont\sansfam=\sevensans\scriptscriptfont
\sansfam=\fivesans
\def\sans{\fam\sansfam\tensans}
\def\Z{{\mathchoice
{\hbox{$\sans\textstyle Z\kern-0.4em Z$}}
{\hbox{$\sans\textstyle Z\kern-0.4em Z$}}
{\hbox{$\sans\scriptstyle Z\kern-0.3em Z$}}
{\hbox{$\sans\scriptscriptstyle Z\kern-0.2em Z$}}}}

\newcount\foot
\foot=1
\def\note#1{\footnote{${}^{\number\foot}$}{\ftn #1}\advance\foot by 1}

\def\frac#1#2{{#1\over #2}}
\def\text#1{\quad{\hbox{#1}}\quad}

\font\ftn=cmr8 scaled\magstephalf

\font\it=cmti10 scaled\magstephalf

\font\titch=cmbx12 scaled\magstep2
\font\titname=cmr10 scaled\magstep2
\font\titit=cmti10 scaled\magstep1
\font\titbf=cmbx10 scaled\magstep2

\nopagenumbers

%%%%%%%%%%%%%%%%%%% title page %%%%%%%%%%%%%%%%%%%%%%%%%%%%%%%%%%
\line{\hfil DFF 236/11/95}
\line{\hfil November 10, 1995}
\vskip2.2cm
\centerline{\titch MAKING QUANTUM GRAVITY CALCULABLE }
%\vskip.5cm
%\centerline{\titch IN QUANTUM GRAVITY}
\vskip1.7cm
\centerline{\titname R. Loll\note{Supported by the European Human
Capital and Mobility program ``Constrained Dynamical Systems"}}
\vskip.5cm
\centerline{\titit Sezione INFN di Firenze}
\vskip.2cm
\centerline{\titit Largo E. Fermi 2}
\vskip.2cm
\centerline{\titit I-50125 Firenze, Italy}
\vskip.3cm
\centerline{and}
\vskip.3cm
\centerline{\titit Max-Planck-Institut f\"ur Gravitationsphysik}
\vskip.2cm
\centerline{\titit Schlaatzweg 1}
\vskip.2cm
\centerline{\titit D-14473 Potsdam, Germany}

\vskip2.0cm
\centerline{\titbf Abstract}
\vskip0.7cm

We describe recent attempts at discretizing canonical quantum gravity
in four dimensions in terms of a connection formulation. This includes
a general introduction, a comparison between the real and complex
connection approach, and a discussion of some open problems.
(Contribution to the proceedings of the workshop ``Recent mathematical
developments in classical and quantum gravity", Sintra, Portugal, July
1995.)

\vfill\eject
\footline={\hss\tenrm\folio\hss}
\pageno=1

%%%%%%%%%%%%%%%%%%%%%%%%%%%%%%%%%%%%%%%%%%%%%%%%%%%%%%%%%%%%%%%%%%%%

\line{\titname 1. Introduction\hfil}

In this article I want to report on some recent developments in
applying discretization methods to quantum gravity in four space-time
dimensions. My emphasis will be on explaining some of the main ideas
and motivations to the non-expert, rather than going much into the
technical details, which for the most part can already be found in
the literature.

As is well known, there is as yet no {\it complete} consistent
framework that would deserve to be called a theory of quantum gravity.
However, there has been a great deal of activity during the last few
years on the canonical quantization of gravity, and most of the
material that follows will be related to these new developments.
Having no very precise idea of what a complete theory of quantum gravity
is eventually going to look like, it is not clear {\it a priori} what
kind of quantities one is interested in calculating (or
approximating). One point of view is that the mathematical structure
of this hypothetical theory would naturally
lead to a set of preferred quantities
that take on a particularly easy form, an idea borne out by recent
results in the canonical approach to quantization.

Our starting point will be the usual Einstein-Hilbert action for
pure gravity,

$$
S^{\rm EH} [{}^{(4)}g] =\int_M d^4 x\sqrt{-{}^{(4)}g} R[{}^{(4)}g],
\eqno(1.1)
$$

\ni possibly with a cosmological constant term. In (1.1),
$g_{\mu\nu}$ is a
non-degenerate, Lorentzian four-metric on the manifold $M$.
The two basic approaches
to the quantization of gravity, as given by the action (1.1), are the
Lagrangian one using a path integral over all possible configurations
of the system, and a Hamiltonian one, based on operator
algebras of preferred functions or
observables. In the first case, one is usually only able to treat
four-metrics with {\it Euclidean} signature,
which leaves one with the problem of
having to continue the results to the Lorentzian regime.
Although the two approaches may look rather different in
concrete implementations, they nevertheless share some
important fundamental questions in common.

In the Lagrangian formulation, one works with functional integrals
which, for example, are of the form

$$
<\hat O>=\int_{\frac{\rm Riem\, M}{\rm Diff\,M}}{\cal D}[g]\,
O([g])e^{-S^{\rm EH}},
\eqno(1.2)
$$

\ni in order to compute vacuum expectation values of
functions $O$ of the four-metric $g_{\mu\nu}$, or rather of
diffeomorphism equivalence classes $[g_{\mu\nu}]$ of metrics,
often called ``geometries". The integration in (1.2) is over
some appropriate set of such geometries, with a suitable measure
${\cal D}[g]$ that is supposed to depend only on equivalence
classes of metrics, i.e., elements of the quotient space
${\rm Riem\, M}/{\rm Diff\,M}$. The usual
strategy of quantum field theory would be to expand (1.2)
perturbatively around the flat Minkowskian (or rather Euclidean)
metric $\eta_{\mu\nu}$, by considering metrics of the form
$g_{\mu\nu}=\eta_{\mu\nu}+ h_{\mu\nu}$ that differ only by a
small amount $h_{\mu\nu}$ from the flat metric. However, for
quantum gravity this strategy fails, since the theory turns out to
be non-renormalizable. One therefore has to try to give meaning
to the integral in some other, non-perturbative way. The
crucial inputs in any such attempt
are the space of quantum configurations to be
integrated over, and the explicit form of the measure ${\cal D}[g]$.

In a Hamiltonian context, the central problem is to identify and
quantize a preferred set of classical variables or observables.
A typical such set consists of a sufficiently
large Poisson algebra of functions which in the quantum theory is
represented as an algebra of operators. As in the Lagrangian case,
one has to determine the space of all quantum configurations and
an inner product on this space in order to be able
to compute scalar products
and matrix elements of observables $O$. For example, in the metric
approach, one seeks to make well-defined an expression like

$$
<\psi([g]),\hat O\phi ([g])>=\int_{\frac{\rm Riem\, \Sigma}
{\rm Diff\,\Sigma}}{\cal D}[g]\psi^* O\phi.\eqno(1.3)
$$

\ni where for simplicity we have assumed that the four-manifold $M$ was
of the form $M=\R\times\Sigma$, with compact spatial slices $\Sigma$.
The approaches to quantum gravity I want to discuss here take the
viewpoint that one should start from discretized versions of
quantities like (1.2) and (1.3), and then obtain the true theory
in an appropriately taken (continuum) limit.

\vskip1.5cm
\line{\titname 2. Lagrangian approaches to discretization\hfil}

Two of the currently active research programs within the path integral
approach are those of Regge calculus and dynamical triangulations.
Quantum Regge calculus has developed from the ground-breaking idea
of Regge [1] to do ``general relativity without coordinates", by
approximating a Riemannian manifold $M$ by a piecewise linear, simplicial
manifold $\cal M$. Since the manifold is piecewise flat, its
curvature is concentrated on lower-dimensional hinges (in $d=4$, on the
two-dimensional triangles).
A simplicial manifold $\cal M$ is characterized
by an incidence matrix which describes the way in which individual
simplices are glued together. In Regge calculus, one fixes an incidence
matrix, i.e. a triangulation, and allows the edge-lengths of the
1-simplices to vary. Edge-length configurations, subject to certain
restrictions, then give rise to (discretized) Riemannian metrics.

In the quantum theory, one tries to approximate the functional
integral by the following product over all 1-simplices,

$$
<\hat O> =\int \prod_i\, d\mu (l_i) O(l) e^{-S^{\rm Regge}(l)},
\eqno(2.1)
$$

\ni where $S^{\rm Regge}$ is a discretized version of the Einstein-Hilbert
action, depending on the link lengths $l_i$ (see, for example, [2]).
One central issue also here is the question of the right measure
$d\mu (l)$. Since many edge length configurations correspond to
the same metric, one expects the appearance of a non-trivial Jacobian
$J$ in $d\mu (l)=\prod_i dl_i J(l)$. In two dimensions, progress
has been reported recently in determining the measure factor
analytically [3], however, no similar results are known in higher
dimensions.  In numerical applications,
the trivial measure $\prod_i dl_i$, or its scale-invariant
version, $\prod_i \frac{dl_i}{l_i}$, have been used, in the hope
that they would lead to reasonable ``effective measures" for the
functional integral.

The second approach, that of ``dynamical triangulations", may be
regarded as a rigid version of Regge calculus. One starts from a set of
equilateral simplices, with all edge lengths set to 1, say, and then
considers all ways in which they may be glued together to obtain
a manifold of a fixed topology (this implies restrictions on the
possible glueings). The main observation then is that distinct such
triangulations $\cal T$ give rise to different metric structures.
One therefore introduces the functional integral

$$
<\hat O>=\sum_{{\cal T},|{\cal T}|=M}\rho ({\cal T})O({\cal T})
e^{-S^{DT}({\cal T})},\eqno(2.2)
$$

\ni where the Einstein-Hilbert Lagrangian (with cosmological constant)
is discretized to $S^{\rm DT}=c_4 N_4 ({\cal T})-c_2 N_2 ({\cal T})$.
$N_2$ and $N_4$ denote the number of 2- and 4-simplices in the
triangulation $\cal T$, and the $c_i$ are bare coupling constants.
This method was first applied in two
dimensions, where a great deal of analytical results are available,
and then generalized to higher dimensions. In this case, hardly any
analytical results exist, and also the numerical analysis is much
harder. The weight factor $\rho ({\cal T})$ is usually set to 1.
Relevant questions in the dynamical triangulations program
are over which (inequivalent) triangulations the
sum (2.2) should be performed, and whether in this way one obtains a
uniform sample of Riemannian geometries (see [4] for recent reviews).

In either approach there are as yet no definite results about the
existence of phase transitions and continuum limits in four
dimensions, and therefore
also the question of the restoration of the diffeomorphism
symmetry (which is not present in the discretized formulation)
in such a limit remains open.

\vskip1.5cm
\line{\titname 3. Discretizing in the canonical picture\hfil}

Let us now turn to the Hamiltonian approach, which will be the
subject of the rest of this article. It will be indispensible for
us to use a connection form $A_a^i$ as our basic variable, instead
of the usual three-metric $g_{ab}$ of the ADM approach. The basic
pair of canonically conjugate variables will consist of an
$SO(3)$-connection $A_a^i(x)$ and a dreibein $\tilde E_i^a$. They
are the exact analogues of the basic variables on the phase space of
a Yang-Mills theory with gauge group $G$, where we would call
the $A_a^i$'s gauge potentials and the $\tilde E_i^a$'s
generalized electric fields. The far-reaching insight that also
canonical gravity might be profitably described by gauge theory-like
variables is due to Ashtekar [5]. The relation with the three-metric
of the ADM canonical variable pair $(g_{ab},\tilde\pi^{ab})$ is given
by $\tilde{\tilde g}^{ab}=\tilde E_i^a\tilde E^{bi}$.
Note that, unlike in Yang-Mills theory, the Einstein-Hilbert action
cannot be rewritten as a functional $S[{}^{(4)}A]$ of a
four-dimensional connection only.  The gauge-theoretic analogy
therefore holds only after the 3+1-decomposition.

In this formulation, and at a kinematical level,
canonical gravity looks like a Yang-Mills
theory with additional constraints, coming from the diffeomorphism
symmetry not present in a gauge theory. Although the basic variables
carry an internal index $i$, associated with the local $SO(3)$-symmetry,
physical quantities, like in Yang-Mills theory, have to be
gauge-invariant. Denoting by ${\cal A}$ the space of all connections
on $\Sigma$, this means that all physically relevant information is
already contained in the quotient space ${\cal A}/{\cal G}$, the
space of gauge equivalence classes of connections.

Another input from Yang-Mills theory is the use of functionals of
$A$ labelled by closed curves $\g$ in the manifold $\Sigma$. These
are the so-called Wilson loops, the traces of path-ordered
exponentials of a connection $A$ along a curve $\g$. Their
importance comes from the fact that they are explicitly gauge-invariant
and form an (over-complete) set of variables on the quotient space
${\cal A}/{\cal G}$. Their non-local nature, that may be regarded
as a defect from a gauge-theoretical point of view, turns out to
be an asset in gravity. This happens because spatial loops are
extended objects with diffeomorphism-invariant properties:
they can link, knot and intersect.

For our purposes, Wilson loops are interesting objects because
they are readily discretized in a lattice framework. Consider a
cubic, three-dimensional $N\times N\times N$-lattice (with
periodic boundary conditions), made up of vertices and
one-dimensional, oriented links or edges $l_i$. The link
holonomy $U_l$ of a link $l$ can be thought of as the
exponentiated integral of the
connection $A$ along that link. Since the $A$'s are non-commuting
matrices in the internal space, one has
to take the {\it path-ordered} exponential of $A$. Note that no
metric structure is necessary to perform the integration
since $A$ is a spatial one-form. We write $U_l(A)={\rm P}\exp
\int_l A$, where one has to keep in mind that $U_l$ takes values
in the gauge group $SO(3)$ and still has
a non-trivial transformation behaviour under gauge transformations
at the beginning and end point of the link $l$. To get rid of this
dependence, one may combine a set of oriented lattice links so that
they form a closed loop $\g$ on the lattice. Taking the trace
of the resulting object, one obtains the {\it lattice Wilson loop}
${\rm Tr}\,U_\g:={\rm Tr}\,U_{l_1}U_{l_2}\dots U_{l_k}$.

\vskip1.5cm
\line{\titname 4. Real vs. complex connection formulation\hfil}

We do not want to derive the Ashtekar variables from scratch,
but rather introduce them in a form that exhibits some
special features. Starting from the Arnowitt-Deser-Misner
variables $(g_{ab},\tilde\pi^{ab})$, one can derive an
equivalent version of canonical gravity based on a pair
$(\tilde P^a_i,K_a^i)$, with an internal $SO(3)$-index $i$,
where $\tilde P_i^a$ is again a dreibein variable, and the
canonically conjugate momentum $K_a^i$ is
closely related to the intrinsic curvature of $\Sigma$.
Compared with the
usual ADM formulation, one obtains three additional first-class
constraints, corresponding to the local $SO(3)$-symmetry.

Consider now the canonical transformation

$$
\tilde E_i^a =-\frac{1}{\b}\tilde P_i^a,
\quad A_a^i=\Gamma_a^i+\b K_a^i,\;
\b\not= 0,\eqno(4.1)
$$

\ni where $\Gamma$ is the $SO(3)$-connection compatible with
the triad $\tilde P_i^a$.
The {\it Ashtekar variables} $A_a^i$ and $\tilde E_i^a$
obey the classical commutation relations

$$
\{A_a^i(x),\tilde E_j^b (y)\}=\,\d_j^i\d_a^b\d^3 (x,y),\eqno(4.2)
$$

\ni since $\Gamma_a^i$ is a function of $\tilde E_i^a$ only.
In this new canonical formulation, the first-class constraints read
(up to constant factors)

$$
\eqalignno{
&\tilde G_i:=\nabla_a\tilde E_i^a =0,&(4.3a)\cr
&\tilde H_a:=F^i_{ab}\tilde E_i^b=0,&(4.3b)\cr
&\tilde{\tilde H}=-\zeta \b^2
  \e^{ijk}\tilde E_i^a\tilde E_j^b F_{ab\,k}
 +2 (\b^2\zeta -1)\tilde E^a_{[i}\tilde E^b_{j]}
 (A_a^i-\Gamma_a^i)(A_b^j-\Gamma_b^j),&(4.3c)}
$$

\ni where $\tilde G_i$ are the three Gauss law constraints
($\nabla=\nabla (A)$ is the covariant derivative with respect to
$A$), the $\tilde H_a$ are the three spatial diffeomorphism
constraints ($F_{ab}^i=\del_a A_b^i-\del_b A_a^i+\e^i{}_{jk}
A_a^j A_b^k$ is the field strength of the connection $A$),
and $\tilde{\tilde H}$ is the Hamiltonian constraint.
The algebraic expression for $\tilde{\tilde H}$ contains two
free parameters, $\b$ and $\zeta$. There are two possible values
for $\zeta$, where

$$
\zeta=\cases{+1 &$\Leftrightarrow\quad$ Euclidean signature;\cr
            -1 &$\Leftrightarrow\quad$ Lorentzian signature.\cr}
$$

\ni For special choices of $\b$ one can
drastically simplify the form of the Hamiltonian $\tilde{\tilde
H}$. In fact, if $(\b^2\zeta-1)=0$, the second term
in (4.3c) drops out and one is left with a {\it polynomial}
Hamiltonian function. This leads to the following choices for
$\b$:

$$
(\b^2\zeta-1)=0 \Leftrightarrow \cases{\b=\pm 1, &Euclidean
  signature;\cr
       \b=\pm i, &Lorentzian signature.\cr}
$$

\ni Our primary interest lies of course with the Lorentzian case.
Nevertheless it is interesting to observe how the difference between
a Lorentzian and a Euclidean signature for the space-time metric
is reflected in the canonical formalism. The derivation for
$\tilde {\tilde H}$ in this compact form is a slightly modified
version of the one given by Barbero [6].
It follows that if (in the Lorentzian case) we want to simplify
the Hamiltonian constraint, we have to use a complex canonical
transformation (i.e. (4.1) with $\b=\pm i$). This is the choice
originally made by Ashtekar [5], and it leads to
a {\it complex} $SO(3)$-connection $A_a^i$ as one of the
basic variables. This gives rise to complications in the sense that
one has to keep track of a set of reality conditions, to ensure
that real, and not complex gravity is described. On the bonus
side, {\it all} of the constraints are now polynomial in
the basic variables $A$ and $\tilde E$. This is important in
the quantization, because polynomiality of a classical phase
space function simplifies the search for a corresponding quantum
operator. (Miraculously, if the real world {\it were} described by
Euclidean metrics, one could achieve this simplification without
ever leaving the real domain.)

The alternative that emerges from this picture (for $\zeta=-1$) is to
set $\b=-1$, say, so that $A_a^i(x)$ is a real $SO(3)$-connection,
and the basic Poisson bracket relations (4.2) have the standard form.
As shown in [6], the Hamiltonian can be rewritten as

$$
\tilde{\tilde H} =\e^{ijk}\tilde E_i^a \tilde E_j^b\,
(F_{ab\,k}(A)-2 R_{ab\,k}(\Gamma))=0,\eqno(4.4)
$$

\ni where $R$ is the curvature of the connection
$\Gamma$, $R_{ab}{}^i=\del_a\Gamma_b^i-\del_b\Gamma_a^i+
\e^i{}_{jk}\Gamma_a^j\Gamma_b^k$. Because of the
$\Gamma$-dependence, $\tilde{\tilde H}$ is non-polynomial in
the basic variables. It is rather similar in structure to the
usual ADM Hamiltonian,

$$
\tilde H^{\rm ADM}[g,\tilde\pi]=-\sqrt{g}\,{}^{(3)}R+\frac{1}{\sqrt{g}}
(\tilde\pi^{ab}\tilde \pi_{ab}-\frac12 \tilde\pi^2)=0,\eqno(4.5)
$$

\ni in that it can be brought into an equivalent
form which is polynomial
{\it modulo powers of the square root of the determinant of
the metric}, $\sqrt{g}=\sqrt{|\tilde{\tilde E}|}$. However, as
is well known, its non-polynomiality is one important reason
that makes the
ADM Hamiltonian $\tilde H^{\rm ADM}$ so hard to quantize,
and has lead to an impasse in the canonical quantization in
the metric approach. So it seems as if nothing had been gained
by using this real connection formulation. We will be returning
to this issue in due course.

Let us summarize the two basic alternatives for the connection
formulation of Lorentzian gravity: either one works with a
simple, polynomial Hamiltonian, leading to complex connection
variables and the need for reality conditions, or one stays
within the real framework, but has to live with a more
complicated, non-polynomial Hamiltonian constraint.

\vskip1.5cm
\line{\titname 5. The lattice discretization\hfil}

Having set the stage for the connection formulation \`a
la Ashtekar in the continuum, we now come back to the
discretized lattice formulation. The basic ingredients are those of
the customary Kogut-Susskind approach of Hamiltonian lattice gauge
theory [7]. For simplicity, we will work with the double cover
$SU(2)$ of $SO(3)$, which in any case is necessary when considering
matter coupling. For the complex case, the gauge group should of
course be $SU(2)^\C=SL(2,\C)$. The lattice analogues of the
variables $(A_a^i,\tilde E_i^a)$ are the link holonomies $U_l$
introduced earlier, and the link momenta $p_i(l)$. We label
vertices of the cubic lattice by a triplet $n=(n_1,n_2,n_3)$,
$n_i=0,1,\dots,N\equiv 0$, and links $l$ by their origin $n$
and a direction $\hat a$. With this notation, the algebra
of the basic quantum operators is given by

$$
\eqalign{
&[\hat U_A{}^B(n,\hat a),\hat U_C{}^D(m,\hat b)]=0\cr
&[\hat  p_i(n,\hat a),\hat U_A{}^C(m,\hat b)]=
-\frac{i}{2}\,\d_{nm}\d_{\hat a\hat b}\, \tau_{iA}{}^B\hat U_B{}^C
(n,\hat a)\cr
&[\hat p_i(n,\hat a),\hat p_j(m,\hat b)]=
 i\, \d_{nm}\d_{\hat a\hat b}\, \e_{ijk}\, \hat p_k(n,
\hat a),}\eqno(5.1)
$$

\ni where the indices $A,B,\dots$ are those of the defining
representation in terms of $2\times 2$-matrices.
The commutators (5.1) are the exact quantization of the
corresponding classical Poisson bracket relations. The holonomy
operators $\hat U_A{}^B$ act by multiplication,
$\hat U_A{}^B\psi (g)=U_A{}^B\psi (g)$, and the link momenta
are represented by the differential operators $\hat p_i=
-i\tau_{iA}{}^B U_B{}^C(\frac{\del}{\del U})_C{}^A$. The $\tau_i$
are the usual Pauli $\sigma$-matrices, rescaled by a factor
$i$, and $g$ stands for a group element (not to be confused
with the same symbol for the metric used elsewhere in this article).

This is the right moment to discuss some of the special features
of lattice {\it gravity}, as opposed to lattice gauge theory.
Firstly, in the real connection formulation, the total lattice
Hilbert space $\cal H$ is given by the direct product over all links
of the square-integrable functions on the group, ${\cal H}=\otimes_l
L^2 (SU(2),dg)$, with $dg$ the Haar measure on $SU(2)$. However,
in the complex $SL(2,\C)$-case, we immediately encounter the
problem that there is no analogue of the left- and right-invariant
Haar measure, because the group is non-compact. Secondly,
we have already mentioned that gravity possesses an extra symmetry,
the spatial diffeomorphisms. This symmetry is badly ``broken"
by the lattice discretization (as it is by any other discretization),
i.e. it is no longer present. Usually this leads one to the
requirement that the symmetry should be restored in the continuum
limit, when all link lengths shrink uniformly to zero, and/or the
number of vertices goes to infinity.

Until recently, no possible resolutions to the scalar product
problem in the complex case were known. (The real case has so far
not been considered.) Some
investigations of lattice versions of Ashtekar gravity were made in
[8-10].  Generally speaking, the lattice computations are rather lengthy,
and not many results have been obtained concerning the algebra of lattice
constraints, both classically and quantum-mechanically (in the latter
case one also has to check for possible anomalies), and the continuum
limit. Renteln was able to show that a particular discretized
version of the quantum algebra of spatial diffeomorphism constraints
reproduces the continuum algebra [9].

What comes to our aid in defining a scalar product for the
$SL(2,\C)$-case is a construction by Hall [11], namely, an integral
transform

$$
\eqalign{
C_t:\,&L^2 (SU(2),dg)\rightarrow L^2(SL(2,\C),d\nu_t)^{\rm
hol}\cr
&[C_t(f)](g_\C):=\int_{SU(2)} dg\, f(g)\rho_t (g^{-1}g_\C)}\eqno(5.2)
$$

\ni where the target space consists only of the {\it holomorphic}
square-integrable functions on the complexified group
((5.2) is to be thought of as a non-linear analogue of the
Segal-Bargmann transform $L^2(\R^n,dx)\rightarrow L^2(\C^n,
d\mu)^{\rm hol}$, where $d\mu$ is the Gaussian measure). The measure
in the holomorphic representation is the heat kernel measure,
depending on a positive, real parameter $t$, and $\rho_t$ is the
heat kernel itself.

These results have been used in [12] to set up a holomorphic lattice
representation. Since the Hall
transform is an isomorphism of Hilbert spaces, it induces a
scalar product on $SL(2,\C)$-lattice wave functions (an explicit
coordinate expression for $d\nu_t$ is not known), and it turns
out that Wilson loop states are indeed square-integrable with
respect to this inner product. The next step then is to find
an appropriate lattice discretization $H^{\rm discr}$ and look
for solutions to the Wheeler-DeWitt equation on the lattice,
$\hat H^{\rm discr}\psi(\g)=0$, and observables $\hat O$
satisfying $[\hat H^{\rm discr},\hat O ]=0$. By $\psi(\g)$ we
denote a generic linear combination of Wilson loop states on
the lattice. If one thinks of taking eventually a continuum
limit by letting the lattice spacing $a$ go to zero, one has to
solve these equations only to lowest order in $a$. In spite of
the finite dimensionality of the lattice discretization, it
is rather non-trivial to make progress on these questions.

What is still needed in this context is a systematic investigation
of all possible discretized Hamiltonians and scalar products.
The choice of the classical $H^{\rm discr}$ is non-unique
because the criterion that $H^{\rm discr}
\buildrel{a\rightarrow 0}\over\longrightarrow H^{\rm cont}$ in
the limit does not fix $H^{\rm discr}$ uniquely. Additional
ambiguities arise through different operator orderings of the
quantum Hamiltonian $\hat H^{\rm discr}$. As for the scalar
product, there are indications that it is too simple to
reflect in a direct way the reality conditions $A+A^\dagger=
2\,\Gamma(\tilde E)$ of the classical phase space formulation.
Rather, it
corresponds to reality conditions of the form $A^\dagger =
A+2 \tilde E$. The implicit metric dependence of this latter
expression could lead to problems with the continuum limit
in lattice gravity. Such a dependence is also present in the
holomorphic continuum formulation [13], where one has
to introduce an auxiliary function that depends on the
lengths (!) of loop segments. Still this does not necessarily
mean that the holomorphic Hilbert space cannot be put to some
use in complex Ashtekar gravity.

For specific choices of the discretized Hamiltonian, some
solutions to the Wheeler-DeWitt equation have been
found. The simplest ones depend on so-called
Polyakov loops [12], which are straight closed lattice loops
winding around the lattice in one of the three directions
(this is possible because of periodicity). More recently,
other solutions have been constructed [14], which are labelled
by multiple plaquette loops (a plaquette on the lattice is
a smallest loop, made up of four links). Although they
solve the Wheeler-DeWitt equation, they are not
square-integrable with respect to the holomorphic scalar
product. There may of course exist other scalar products
in which those solutions have finite norm.

\vskip1.5cm
\line{\titname 6. Computing volumes\hfil}

Apart from the issue of the scalar product, there is another
reason why the solutions to the discretized Wheeler-DeWitt
equation found so far are probably not
very interesting physically. In order to understand this,
we need to study the so-called {\it volume operator}.
It comes from the classical volume function,

$$
{\cal V}({\cal R})=\int_{\cal R} d^3x\;\sqrt{g}=
\int_{\cal R} d^3x\;\sqrt{\frac{1}{3!} |\e_{abc}\,\e^{ijk}
\tilde E^a_i \tilde E^b_j \tilde E^c_k |},\eqno(6.1)
$$

\ni which measures the volume of a spatial region ${\cal R}
\subset\Sigma$. Note that $\cal V$ is not an observable in pure
gravity in the sense that it does not Poisson-commute with
the Hamiltonian constraint $\tilde{\tilde H}$). It however
becomes an observable if the region $\cal R$ is defined
intrinsically in some way by the presence of matter fields [15].
This quantity can be studied also on the lattice, because
it can be discretized straightforwardly and written as a
function of the lattice momenta,

$$
{\cal V}^{\rm latt}=\sum_{n\in{\cal R}} \sqrt{\frac{1}{3!}
|\e_{abc}\,\e^{ijk}\,
p_i(n,\hat a) p_j(n,\hat b) p_k(n,\hat c) |},\eqno(6.2)
$$

\ni where the integral has been substituted by a sum over all
vertices lying within a lattice region $\cal R$.
Because of the presence of both the modulus and the
square root, it is not {\it a priori} obvious how to
quantize (6.2). However, in this case we are lucky, because
both in the real and the holomorphic representation the
momenta $\hat p$ are selfadjoint operators [16]. One can
therefore go to a basis where the terms cubic in $\hat p$
under the modulus are diagonal operators, and the
modulus and square root can be defined in terms of their
eigenvalues.

In order to diagonalize the volume operator, it is very
convenient to use a particular set of gauge-invariant states
in the loop representation, the (generalized) {\it spin
network states}, as was first pointed out in [15]. A spin
network state may be represented as a certain antisymmetrized
linear combination of Wilson loop states. In order to
describe a spin network on the lattice, one needs to
associate an ``occupation number" $j(l)\in\Z^+$ with each
lattice link (subject to a number of restrictions). This
can be represented by drawing $j(l)$ lines through the link
$l$. To obtain a multiple Wilson loop, one has to connect
these lines at the lattice vertices in a gauge-invariant
manner. The contractors at the vertices are the second
ingredient in the definition of a spin network state.
Details of this construction can be found in [15-17].

It can be shown that the spin network states span the space
of all gauge-invariant lattice functions [18,17]. The operator
$\hat {\cal V}^{\rm latt}$ is ``almost diagonal" in
terms of the spin networks, i.e. the diagonalization is
reduced to diagonalizations within finite-dimensional
eigenspaces of $\hat {\cal V}^{\rm latt}$, and its spectrum
is discrete [15]. In determining
the eigenvalues of the volume operator explicitly, it suffices to
study the action of the operator $\hat D(n):=\e_{abc}\,\e^{ijk}\,
\hat p_i(n,\hat a) \hat p_j(n,\hat b) \hat p_k(n,\hat c)$
around a vertex $n$. Moreover it is useful to consider
separately cases of different valence, where the valence
counts the number of links with non-vanishing $j(l)$ meeting
at $n$.

Because of the antisymmetry of the operator $\hat D(n)$, it annihilates
spin networks which are two-valent at $n$. Although it is not
immediately obvious, one can construct a general argument
that also trivalent vertices are annihilated, i.e.
spin network states that possess at most trivalent
intersections have ``zero volume" [16]. This is the reason
why we considered the solutions for the Hamiltonian constraint
equation described above as probably uninteresting. One obtains the
first non-trivial eigenvalues for four-valent intersections.
Some explicit spectral formulae for such spin networks were
derived in [19]. One technical problem one encounters in the
diagonalization is the occurrence of linearly dependent
spin network states (because not all ways of contracting
flux lines at vertices are independent). These states have
to be eliminated to avoid spurious eigenvalues.

\vskip1.5cm
\line{\titname 7. Outlook, and a new perspective\hfil}

Apart from yielding interesting information about the volume
operator itself, the construction described in the previous
section also throws new light on the {\it real} connection
formulation. We mentioned earlier that Barbero's Hamiltonian
can be written in polynomial form up to powers of $\sqrt{g}$.
Now observe that in the diagonalized spin network basis,
where the volume operator is diagonal, also the determinant
$g$ (which may be written as a cubic polynomial
in the momenta) is diagonal. This means that in this basis
it is straightforward to define all operators of the kind
$\hat{\sqrt{g}^k}$. It turns out that this enables one to quantize
a discretized version of the Hamiltonian (4.4) on the lattice [20].

Of course this quantum Hamiltonian is still more complicated
than the one of the complex Ashtekar formulation,
$\tilde{\tilde H} =\e^{ijk}\tilde E_i^a \tilde E_j^b F_{ab\,k}$,
in that it has an additional potential term. On the other
hand, there are no problems with the choice of a scalar
product which is simply given by the ordinary Haar measure on
$SU(2)$, as in lattice Yang-Mills theory. However, contrary to the
ADM formulation, there is now an obvious way to quantize
arbitrary functions of the determinant of the metric.

If one wants to pursue this line of thought further, and try
to solve the Wheeler-DeWitt equation, say, it is clear that
one a) needs an efficient general method for generating
volume eigenstates, and b) must eliminate the zero-volume
states because of the inverse powers of the metric occurring
in the Barbero Hamiltonian. Thus, unlike in complex Ashtekar
gravity, in the real context one cannot extrapolate smoothly
to the case where the (quantized) three-metric is allowed to be
degenerate.

To summarize, there has been progress in Hamiltonian lattice
gravity, although many details remain to be worked out.
Crucial in all recent developments has been the use of connection
variables, and the associated gauge-invariant Wilson loops.
There are two basic alternatives to proceed in the lattice
framework. The first one is that corresponding to complex $A$;
there it remains to find solutions ``with volume"
to $\hat H\psi=0$ (although it is not strictly necessary
to work in terms of spin network states),
and to investigate different Hamiltonians and scalar products.
The second way is that based on a real connection $A$; as we have
seen, in this case the use of a diagonalized spin network basis is
mandatory; the search for solutions to the Wheeler-DeWitt
equation has only just begun [20]. Once some solutions have
been found, it would be very interesting to compare the
results of both the real and the complex formulation, and to
see whether they are equivalent. Other issues that remain
to be addressed are whether and how the continuum limit
should be taken, and whether and how diffeomorphism invariance
should be incorporated in the lattice picture. Also it would
be most interesting to try to study matter couplings in the
same framework.

\ni{\it Acknowledgement.} I am grateful to the organizers of
the Sintra meeting for their hospitality, as well as
to the members of the
Max-Planck-Institut where these notes were written up.

\vskip2cm
\vfill\eject

\line{\titname References\hfil}

\item{[1]} Regge, T.: General relativity without coordinates,
 {\it Nuovo Cim.} 19 (1961) 558-71

\item{[2]} Williams, R.M. and Tuckey, P.A.: Regge calculus: A
  bibliography and brief review, {\it Class. Quant. Grav.} 9
  (1992) 1409-22; Fr\"ohlich, J.: Regge calculus and discretized
  functional integrals, in {\it Non-perturbative quantum field theory}:
  selected papers of J\"urg Fr\"ohlich, World Scientific (Singapore)
  1992, 523-45

\item{[3]} Menotti, P. and Peirano, P.P.: Faddeev-Popov
  determinant in 2-dimensional Regge gravity, {\it Phys.
  Lett.} 353B (1995) 444-9

\item{[4]} Ambj\o rn, J., Jurkiewicz, J. and Watabiki, Y.:
  Dynamical triangulations, a gateway to quantum gravity?,
  {\it preprint} Bohr Inst., NBI-HE-95-08, e-print archive:
  hep-th 9503108;
  Jurkiewicz, J.: Simplicial gravity and random surfaces,
  {\it Nucl. Phys.} B {\it Proc. Suppl.} 30 (1993) 108-21

\item{[5]} Ashtekar, A.: New variables for classical and quantum
  gravity, {\it Phys. Rev. Lett.} 57 (1986) 2244-7; A new
  Hamiltonian formulation of general relativity, {\it Phys.
  Rev.} D36 (1987) 1587-603

\item{[6]} Barbero, J.F.: Real Ashtekar variables for Lorentzian
  signature space-times, {\it Phys. Rev.} D51 (1995) 5507-10

\item{[7]} Kogut, J. and Susskind, L.: Hamiltonian formulation of
  Wilson's lattice gauge theories, {\it Phys. Rev.} D11 (1975)
  395-408; Kogut, J.B.: The lattice gauge theory approach
  to quantum chromodynamics, {\it Rev. Mod. Phys.} 55 (1983) 775-836

\item{[8]} Renteln, P. and Smolin, L.: A lattice approach to spinorial
  quantum gravity, {\it Class. Quant. Grav.} 6 (1989) 275-94

\item{[9]} Renteln, P.: Some results of SU(2) spinorial lattice
  gravity, {\it Class. Quant. Grav.} 7 (1990) 493-502

\item{[10]} Bostr\"om, O., Miller, M. and Smolin, L.: A new
  discretization of classical and quantum general relativity, {\it
  preprint} G\"oteborg U. ITP 94-5 and Syracuse U. SU-GP-93-4-1

\item{[11]} Hall, B.C.: The Segal-Bargmann coherent state transform
  for compact Lie groups, {\it Journ. Funct. Anal.} 122 (1994) 103-51

\item{[12]} Loll, R.: Non-perturbative solutions for lattice quantum
  gravity, {\it Nucl. Phys.} B444 (1995) 619-39

\item{[13]} Ashtekar, A., Lewandowski, L., Marolf, D., Mour\~ao, J.
  and Thiemann, T.: Coherent state transform for spaces of connections,
  {\it preprint} Penn State U., Dec 1994, e-print archive: gr-qc
  9412014

\item{[14]} Ezawa, K.: Multi-plaquette solutions for discretized
  Ashtekar gravity, {\it preprint} Osaka U. OU-HET/223, e-print
  archive: gr-qc 9510019

\item{[15]} Rovelli, C. and Smolin, L.: Discreteness of area and
  volume in quantum gravity, {\it Nucl. Phys.} B442 (1995)
  593-619

\item{[16]} Loll, R.: The volume operator in discretized quantum
  gravity, {\it Phys. Rev. Lett.} 75 (1995) 3048-51

\item{[17]} Rovelli, C. and Smolin, L.: Spin networks and quantum
  gravity, {\it preprint} Penn State U. CGPG-95/4-4, e-print
  archive: gr-qc 9505006

\item{[18]} Baez, J.B.: Spin network states in gauge theory, to
  appear in {\it Adv. Math.}, e-print archive: gr-qc 9411007

\item{[19]} Loll, R.: Spectrum of the volume operator in quantum
  gravity, {\it preprint} INFN Firenze, DFF 235/11/95, e-print
  archive: gr-qc 9511030

\item{[20]} Loll, R., {\it preprint} INFN Firenze, in preparation

\end